\documentclass[pre,twocolumn,amsmath,groupedaddress,floatfix]{revtex4}

\usepackage{graphicx}
\begin{document}                                                                                   
       
\title{Ion Concentration Dynamics as a Mechanism for Neuronal Bursting} 

\author{Ernest Barreto}
\email[Email address: ]{ebarreto@gmu.edu}

\author{John R. Cressman} 
\email[Email address: ]{jcressma@gmu.edu}
\affiliation{Center for Neural Dynamics, Department of Physics \& Astronomy, and\\
The Krasnow Institute for Advanced Study, George Mason University, Fairfax, Virginia 22030 USA}

\begin{abstract}
We describe a simple conductance-based model neuron that includes intra- and extra-cellular
ion concentration dynamics and show that this model exhibits periodic bursting.
The bursting arises as the fast spiking behavior of the neuron is modulated by
the slow oscillatory behavior in the ion concentration variables, and vice versa.
By separating these time scales and studying the bifurcation structure of the neuron,
we catalog several qualitatively different bursting profiles that are strikingly similar
to those seen in experimental preparations.
Our work suggests that ion concentration dynamics may play an important role in
modulating neuronal excitability in real biological systems.
\\
\\
This is an author-generated version of an article that appears in the Journal of Biological Physics {\bf 37}, 361-373 (2011). Please use that citation. The official publication is available at http://www.springerlink.com/content/v52215p195159211 (DOI: 10.1007/s10867-010-9212-6).
\end{abstract}

\maketitle

\section{Introduction}

The Hodgkin-Huxley equations \cite{HH} are of fundamental importance in theoretical neuroscience. These
equations assume that the intra- and extra-cellular ion concentrations of sodium and potassium are constant.
While this may be a reasonable assumption for the squid giant axon preparation (for which the equations were
originally developed), its validity in other cases is not clear. In the mammalian brain, for example,
the neurons are much smaller and they are more tightly packed, resulting in significantly smaller intra-
and extra-cellular volumes. Thus, typical ionic currents can have a much larger
effect on the ion concentrations in this case.

The effects of extracellular potassium ($[K]_o$) accumulation on neuronal excitability have long been
recognized \cite{FH,Grafstein,Green,FR}, and deficiencies in $[K]_o$ regulation have been implicated in
various types of epilepsy (for a review, see \cite{FrohlichReview}) and spreading depression
\cite{Kager2002,Somjen2008b}. More recently, computational studies
have begun to clarify the role of impaired $[K]_o$ regulation
\cite{Bazhenov2004,Frohlich2006,Park,Kager2007,Somjen2008a,Postnov2009,Frohlich2010}
as well as other varying ion concentrations \cite{Kager2000,SomjenBook}.

In this work, we consider from a dynamical systems perspective the role of ion concentration dynamics
in the generation of periodic bursting behavior. To emphasize the generality of our approach, we
base our model on the standard Hodgkin-Huxley equations. We augment these with additional equations
that describe the dynamics of both intra- and extra-cellular sodium and potassium. The
inclusion of sodium is relatively novel and plays a crucial role in the dynamics described here.
We also include terms describing pumps, extracellular diffusion, and a simple glial buffering system.
A different analysis of this system was presented in \cite{CressmanJCNS}.

\section{Model}
We begin by explicitly adopting the standard convention that an outward membrane current is defined
as being positive \cite{DayanAbbott}. Thus, the membrane potential $V$ is given by
 \begin{equation}
 {C}\frac{{dV}}{{dt}} =  - {I_{membrane}}
 \label{Veqn}
 \end{equation}
where $I_{membrane}$ represents the sum of the various membrane currents. 
We aim in this work to consider a very simple and general model neuron. Hence we include only
the standard Hodgkin-Huxley sodium current (with instantaneous activation), the delayed-rectifier potassium current,
and leak current. We write the latter in terms of separate sodium, potassium, and chloride contributions \cite{{Kager2000}}.
Thus,
\begin{equation}
\begin{array}{l}
 I_{membrane} =  {I_{Na}} + {I_K} + {I_{Cl}} \\ 
 {I_{Na}} = {g_{Na}}{\left[ {{m_\infty }} \right]^3}h\left( {V - {E_{Na}}} \right) + {g_{NaL}}\left( {V - {E_{Na}}} \right) \\ 
 {I_K} =  {g_K}{n^4}  \left( {V - {E_K}} \right) + {g_{KL}}\left( {V - {E_K}} \right) \\ 
 {I_{Cl}} = {g_{ClL}}\left( {V - {E_{Cl}}} \right),
 \end{array}
 \label{currents}
\end{equation}
where the $g_i$ ($i=$ Na, K, Cl) are maximum conductances.
Time is measured in milliseconds, voltage in millivolts, and ${C},I,$ and $g$ are measured in units per
unit of membrane area, i.e., $\mu F/cm^2$, $\mu A/cm^2$, and $mS/cm^2$, respectively.
The reversal potentials $E_i$ are given in terms of the instantaneous intra- and extracellular ion concentrations
by Nernst equations:
\begin{equation}
\begin{array}{l}
 {E_{Na}} = 26.64\ln \left( {\frac{{{{[Na]}_o}}}{{{{[Na]}_i}}}} \right) \\ 
 {E_K} = 26.64\ln \left( {\frac{{{{[K]}_o}}}{{{{[K]}_i}}}} \right).
 \end{array}
 \label{Nernst}
\end{equation}
We fix ${E_{Cl}} =   - 81.9386\;{\rm{mV}}$.

The extracellular potassium and intracellular sodium concentration dynamics are given by
\begin{equation}
\begin{array}{l}
 \tau \frac{{d{{[K]}_o}}}{{dt}} =  {\gamma \beta {I_K} - 2\beta {{\tilde I}_{pump}} - {{\tilde I}_{glia}} - {{\tilde I}_{diffusion}}}  \\ 
 \tau \frac{{d{{[Na]}_i}}}{{dt}} =  { - \gamma {I_{Na}} - 3{{\tilde I}_{pump}}},
 \end{array}
 \label{concdiffeqs}
\end{equation}
where the concentrations are measured in millimolar (mM). $\gamma=4.45 \times {10^{ - 2}}$ is a unit conversion factor that converts the membrane
currents into mM/sec (see Appendix), and $\beta=7$ is the ratio of the intracellular to extracellular volume \cite{SomjenBook}. The terms with tildes
are the molar currents (mM/sec) given below, and $\tau=10^3$ balances the time units.

The pump, glia, and diffusion molar currents are given by
\begin{equation}
\begin{array}{l}
 {{\tilde I}_{pump}} = \rho {\left( {1 + \exp \left( {\frac{{25 - {{[Na]}_i}}}{3}} \right)} \right)^{ - 1}}\left( {\frac{1}{{1 + \exp \left( {5.5 - {{[K]}_o}} \right)}}} \right) \\ 
 {{\tilde I}_{glia}} = {G}{\left( {1 + \exp \left( {\frac{{18 - {{[K]}_o}}}{{2.5}}} \right)} \right)^{ - 1}} \\ 
 {{\tilde I}_{diffusion}} = \varepsilon \left( {{{[K]}_o} - {k_{bath}}} \right).
 \end{array}
 \label{pumpgliadiffusion}
\end{equation}
We set the default parameter values to $\rho  = 1.25\;{\rm{mM/sec}}$, $G = 66.666\;{\rm{mM/sec}}$, and $\varepsilon  = 1.333\;{\rm{Hz}}$.
${k_{bath}}$ represents the potassium concentration in the reservoir, i.e., the bathing solution for a slice preparation, or the vasculature {\it in vivo}.
We set ${k_{bath}}= 4\;{\rm{mM}}$ for normal physiological conditions.

The intracellular potassium and extracellular sodium concentrations are obtained from the following simplifying
assumptions that allow us to reduce the dimensionality of the system \cite{CressmanJCNS}:
\begin{equation}
\begin{array}{l}
 {[K]_i} = 140{\rm{mM}} + \left( {18{\rm{mM}} - {{[Na]}_i}} \right) \\ 
 {[Na]_o} = 144{\rm{mM}} - \beta \left( {{{[Na]}_i} - 18{\rm{mM}}} \right).
 \end{array}
 \label{assumptions}
\end{equation}
The first assumption is that the sodium membrane current is the dominant means by
which sodium is transported across the membrane, and that during the course of an
action potential, the transport of sodium and potassium are simply related. The
second assumption is that the total amount of sodium is conserved.

The remaining parameters and the equations for the gating variables are given
in the Appendix.

\section{Results}

\subsection{Fixed ion concentrations}
We begin with a discussion of the dynamical structure of our model subject to
constant values of the ion concentrations. That is, we set $[K]_o$ and $[Na]_i$ to fixed
predetermined values, and obtain the remaining concentrations using Equations
(\ref{assumptions}). We then examine the behavior of the neuron as given by
Equations (\ref{Veqn})--(\ref{Nernst}).

\begin{figure}
\begin{center}
\scalebox{0.8}{\includegraphics{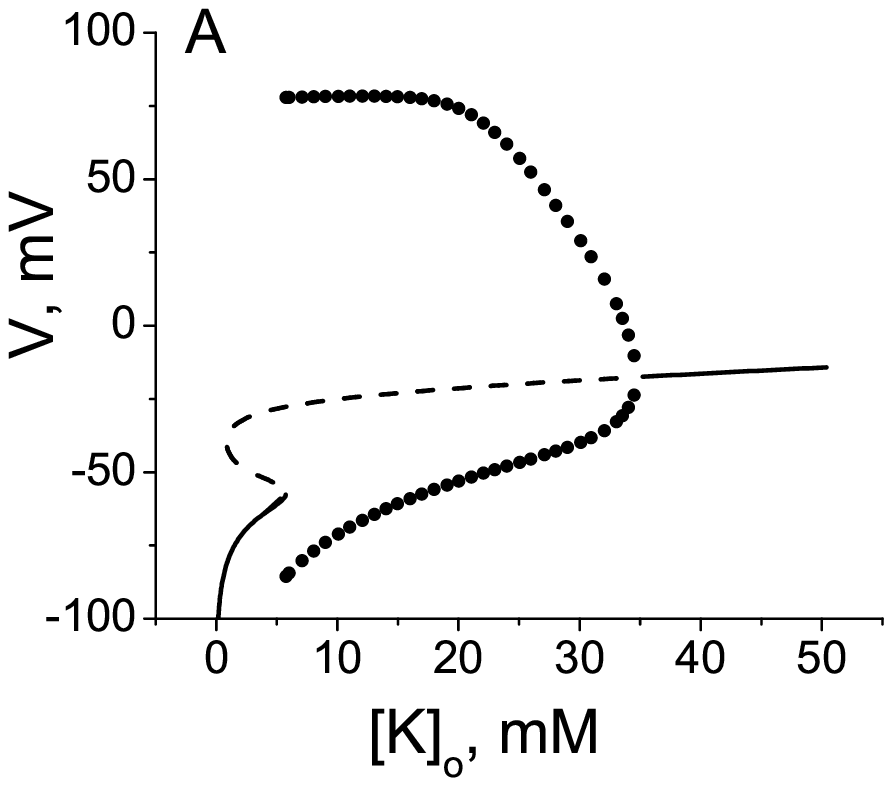}}
\scalebox{0.8}{\includegraphics{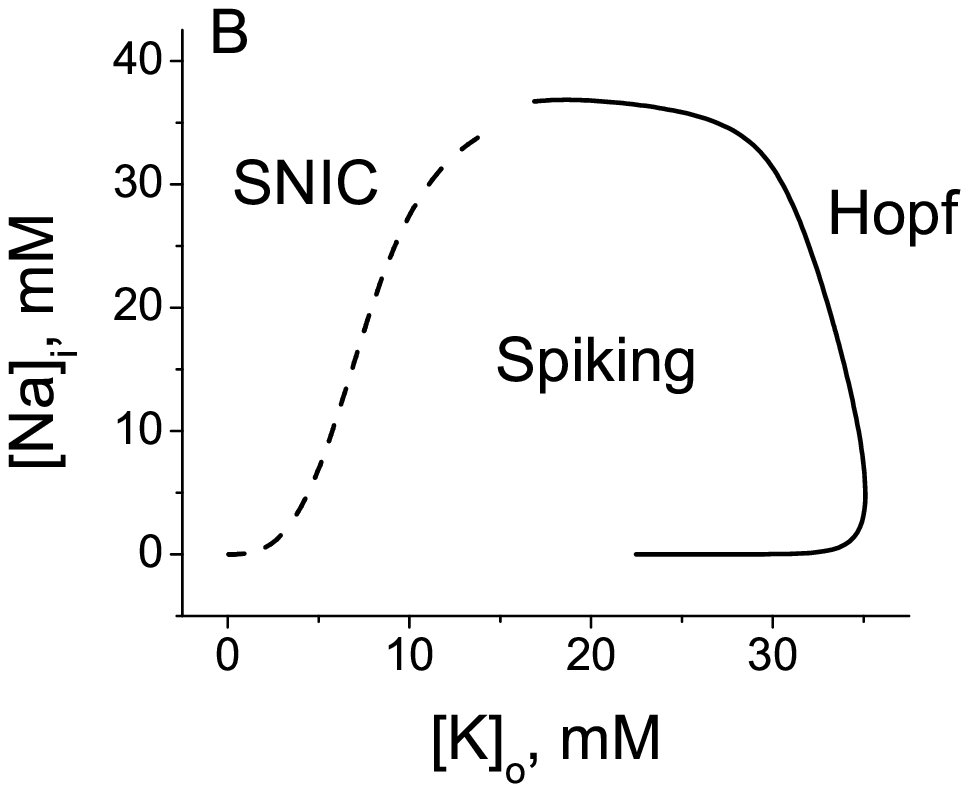}}
\caption{Bifurcation diagrams describing the neuron's asymptotic behavior with fixed ion concentrations. In (A), $[Na]_i=10$mM.
The detailed structure in the upper part of (B) is shown in Figure \ref{UpperPart2Dbif}.}
\label{SNICHOPF}
\end{center}
\end{figure}

Figure \ref{SNICHOPF} shows bifurcation diagrams
obtained under these conditions. The features in these diagrams clarify the neuron's
bursting behavior as the ion concentrations undergo slow oscillations, as explained below.

Figure \ref{SNICHOPF}A is constructed by holding $[Na]_i$ fixed at $10$mM and plotting
the asymptotic values of the membrane potential $V$ versus several fixed values of $[K]_o$.
Below approximately $5.7$mM, the neuron is attracted to a stable equilibrium
(shown as a solid line) that corresponds to the resting state. At approximately $[K]_o=5.7$mM,
this equilibrium coalesces with a coexisting unstable equilibrium (dotted line) in
a saddle-node bifurcation that occurs on an invariant closed curve (of infinite period) that
appears simultaneously. This scenario is known as a SNIC bifurcation \footnote{SNIC stands
for \underline{S}addle-\underline{N}ode on \underline{I}nvariant \underline{C}ircle (see, e.g., \cite{Eugenebook}).
This codimension-one bifurcation has also been called a SNIPER bifurcation, for \underline{S}addle-\underline{N}ode
\underline{I}nfinite-\underline{PER}iod \cite{SNIPERref}.}. Beyond this,
for values of $[K]_o$ between $5.7$mM and $35.2$mM, a stable limit cycle appears,
reflecting regular spiking in the neuron. This is depicted in the diagram by filled
circles that mark the maximum and minimum values of the membrane voltage during a cycle. For
increasing values of $[K]_o$ approaching $[K]_o=35.2$mM from below, the amplitude of this periodic
orbit decreases and the orbit eventually merges with the coexisting unstable equilibrium in a supercritical
Hopf bifurcation. For $[K]_o>35.2$mM, a stable equilibrium is found -- this is the state of
depolarization block \cite{Bikson}.

Figure \ref{SNICHOPF}B is a two-dimensional bifurcation diagram which shows the location of the SNIC and Hopf
bifurcations as the value of $[Na]_i$ is varied. These curves delineate the boundaries of different attracting
behaviors of the neuron. To the left of the SNIC curve, the neuron is attracted to the resting
equilibrium. Between the SNIC and the Hopf curves, the neuron exhibits regular spiking, and to
the right of the Hopf curve, the neuron is attracted to the depolarization block equilibrium.
(The detailed structure at the top of this diagram is discussed below.)

\subsection{Dynamic ion concentrations}
\begin{figure}
\begin{center}
\scalebox{0.9}{\includegraphics{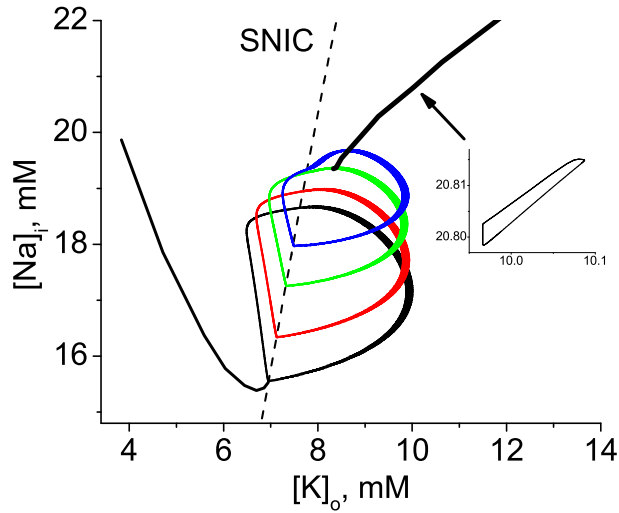}}
\caption{(Color) Asymptotic behavior of the ion concentrations as $k_{bath}$ (in mM) is varied. The curve on the
left denotes stable equilibria and corresponds to $k_{bath}$ ranging from $4.0$ to $7.615$. Projections of limit
cycles (loops) are shown for $k_{bath}=7.62, 8.0, 8.5$, and $8.95$ mM; the thickness of the loops on the right
reflect small fast ion concentration fluctuations due to spiking behavior.
The curve in the upper right, for $k_{bath}$ ranging from $8.8$ to $15.0$,
denotes small-amplitude loops that correspond to tonic spiking (see inset).}
\label{DriftingLoops}
\end{center}
\end{figure}

We now describe the behavior of the full system, in which the ion concentrations are allowed
to evolve dynamically. In Figure \ref{DriftingLoops}, we plot the asymptotic behavior of the ion concentrations for
several values of $k_{bath}$. Also included in the figure is a portion of the SNIC curve from Figure \ref{SNICHOPF}B.

At the default parameter values described above (with $k_{bath}=4.0$ mM), the
entire system approaches a stable equilibrium resting state for which the membrane voltage and
the ion concentrations assume fixed values. As $k_{bath}$ is increased, these equilibrium values
change and sweep out the solid curve shown on the left of Figure \ref{DriftingLoops}.
At approximately $k_{bath}=7.615$ mM, this curve collides with the SNIC
boundary. Just beyond this value, the system jumps to a limit cycle. As $k_{bath}$
continues to increase, the projection of this limit cycle onto the ion concentration
variables drifts upward and to the right, as shown in the figure. Henceforth,
for brevity, we refer to such limit cycle projections as ``loops".

Note that these large-amplitude loops straddle the SNIC curve. In addition, they
have periods on the order of several tens of seconds. Consequently, as the system alternately
transitions between the resting and the spiking regions, there is ample time to exhibit those
asymptotic behaviors. That is, the neuron bursts. For example, for $k_{bath}=8.0$ mM, the ion
concentrations are attracted to the second (red) loop in Figure \ref{DriftingLoops}, and
the corresponding behavior of the membrane voltage is shown below in Figure
\ref{Burst_types}A.

At approximately $k_{bath}=9.0$ mM, the large-amplitude loop disappears. For larger $k_{bath}$,
the ion concentrations exhibit very small-amplitude loops, as shown in the inset of
Figure \ref{DriftingLoops}. Since this loop lies entirely within the spiking region, the neuron
exhibits tonic spiking. Indeed, the loop itself represents the small changes in the ion concentrations
due to individual action potentials.

We note in passing that we have observed multistability \cite{Cymbalyuk1,Cymbalyuk2,FrohlichBazhenov2006}.
For values of $k_{bath}$ approximately in
the interval $(8.8, 9.0)$ mM, a bursting solution coexists with a tonically-spiking state for the
same parameter values. This is consistent with the analysis based on a reduced model in \cite{CressmanJCNS}.

\subsection{Catalog of bursting types}
The results presented above demonstrate that for parameter values in appropriately chosen ranges,
the system evolves on a limit cycle whose projection onto the ion concentration variables
forms a loop that straddles the SNIC curve, and the neuron bursts.
Bursting behaviors of various qualitatively different kinds
can be exhibited by the system if similar ion concentration loops straddle the bifurcation
curves of Figure \ref{SNICHOPF}B in different ways. Accordingly, we can catalog all the
possible arrangements, and examine the nature of the resulting bursting patterns.

\begin{figure}
\begin{center}
\scalebox{0.9}{\includegraphics{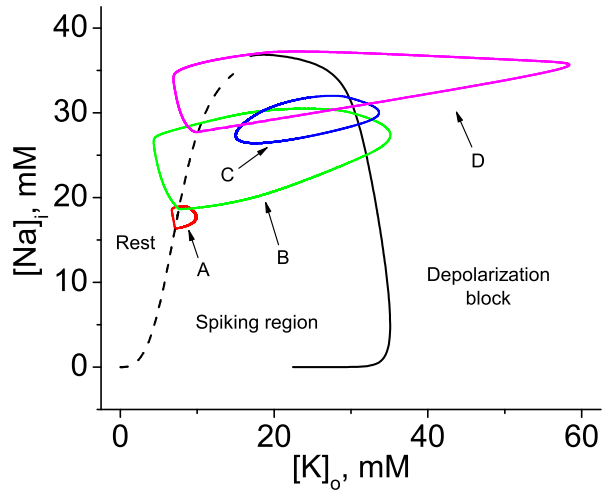}}
\caption{(Color) Loops A-D represent the time evolution of the ion concentrations
as the system exhibits limit cycle behavior. The loops are traversed in a counter-clockwise manner.
The dotted and solid lines are the SNIC and Hopf bifurcation curves, respectively.}
\label{Burst_loopmap}
\end{center}
\end{figure}

\begin{figure*}
\begin{center}
\scalebox{0.8}{\includegraphics{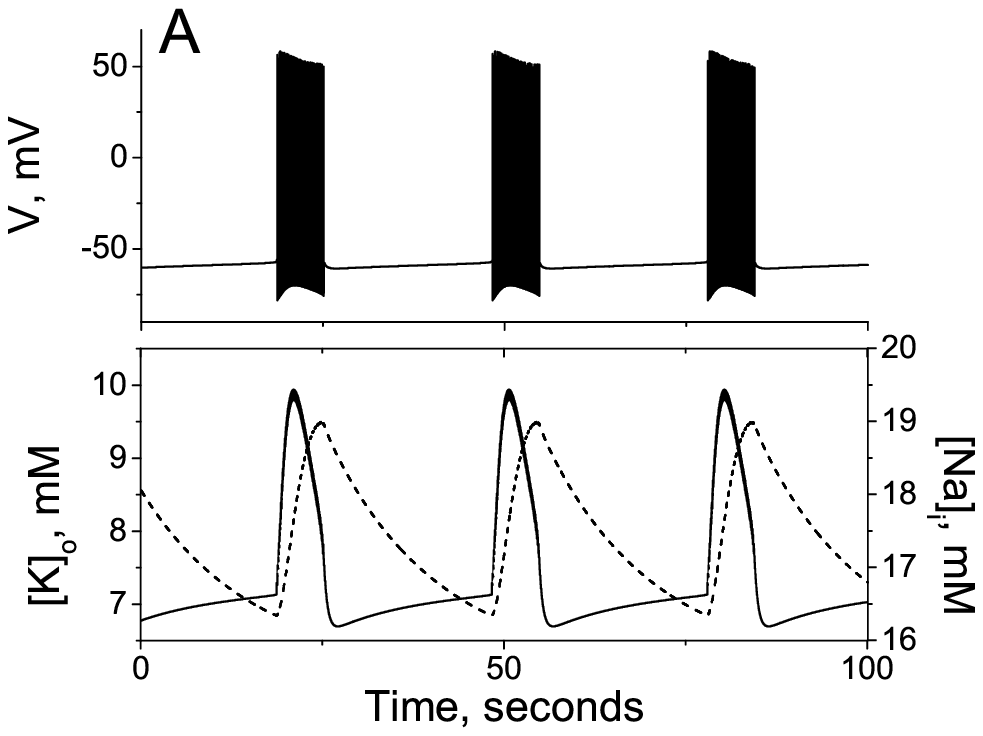}}\scalebox{0.8}{\includegraphics{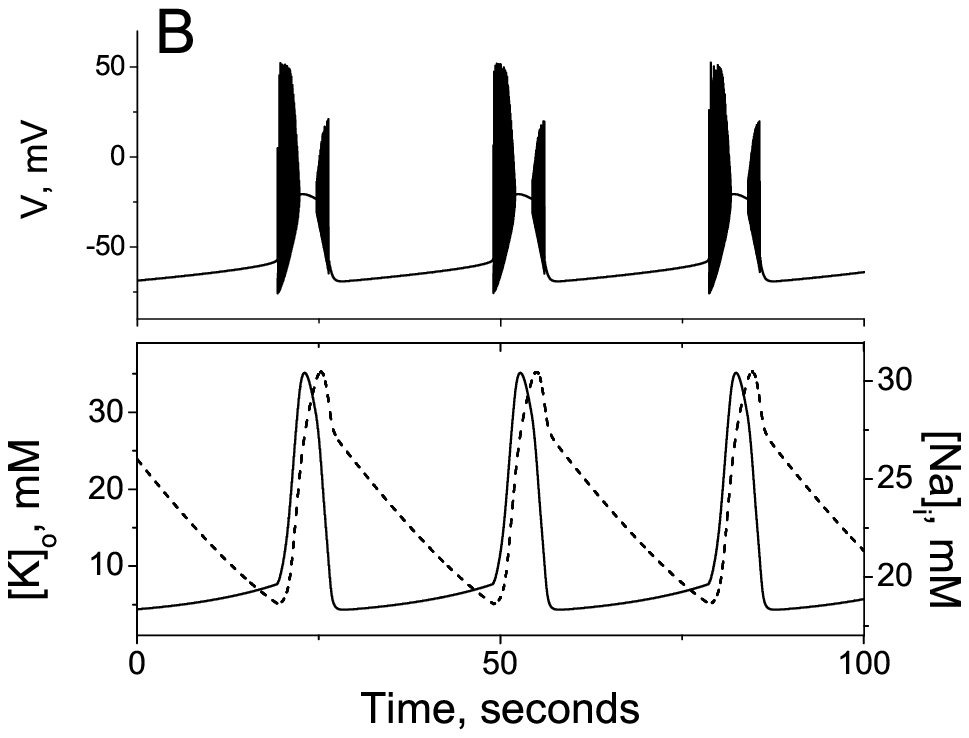}}
\scalebox{0.8}{\includegraphics{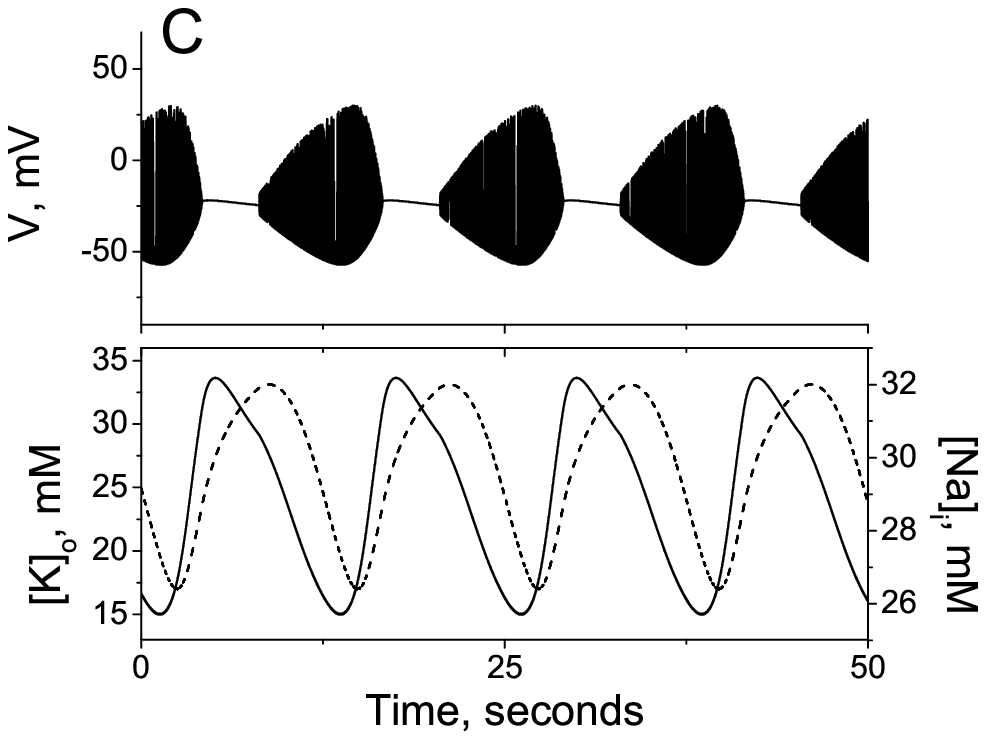}}\scalebox{0.8}{\includegraphics{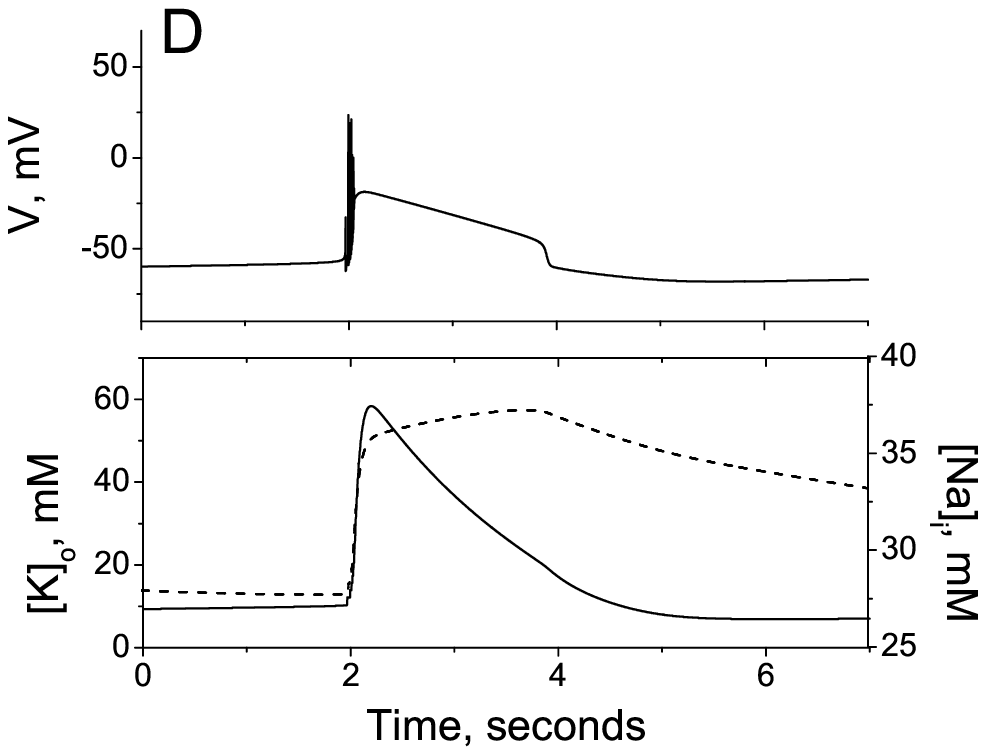}}
\caption{Four qualitatively different bursting patterns corresponding to the four loops shown in Figure \ref{Burst_loopmap}.
In the lower panels, solid curves represent $[K]_o$ (left vertical axes), and dotted curves represent $[Na]_i$ (right vertical axes).
The event in (D) recurs with a period of approximately $16.5$ seconds, but for clarity, only a portion of one such event is shown.}
\label{Burst_types}
\end{center}
\end{figure*}

Figure \ref{Burst_loopmap}
shows four different ion concentration loops labeled A, B, C, and
D. Loop A is the $k_{bath}=8.0$mM loop discussed above, and the corresponding membrane voltage
trace is shown in Figure \ref{Burst_types}A.
Loop B ($G=20$, $\epsilon=0.133$, and $k_{bath}=22$) straddles both the SNIC and the Hopf
bifurcation curves, and hence displays a qualitatively different bursting pattern. As the
loop is traversed, the ion concentration trajectory
moves from the resting region into the spiking region by crossing the SNIC curve,
and then continues across the Hopf curve into the depolarization block region. It then bends
around and crosses these regions in reverse order, and the cycle repeats. Consequently, the
membrane potential shows a bursting pattern that moves from quiescence to spiking to
depolarization block and back again, as is shown in Figure \ref{Burst_types}B.
Loop C ($G=6$, $\epsilon=0.7$, and $k_{bath}=22$)
straddles only the Hopf curve, and hence the membrane potential displays round-shaped bursts,
reflecting the supercritical nature of the Hopf bifurcation, as shown in Figure (\ref{Burst_types}C).
Finally, Loop D ($\rho=0.9$, $G=10$, $\epsilon=0.5$, $k_{bath}=20$, and $\gamma=1.0$)
is similar to loop B, but by examining the membrane voltage in Figure \ref{Burst_types}D,
one sees that the event termination transitions smoothly from depolarization block
back to the resting level without exhibiting any spikes.  This is because the return trip along
the upper portion of the loop avoids the Hopf bifurcation. We clarify this transition in the next
section. (Note that, for nearby parameter sets, it is possible to observe bursts of this type
with more spikes at the event onset than are shown here.)

\begin{figure}
\begin{center}
\scalebox{0.9}{\includegraphics{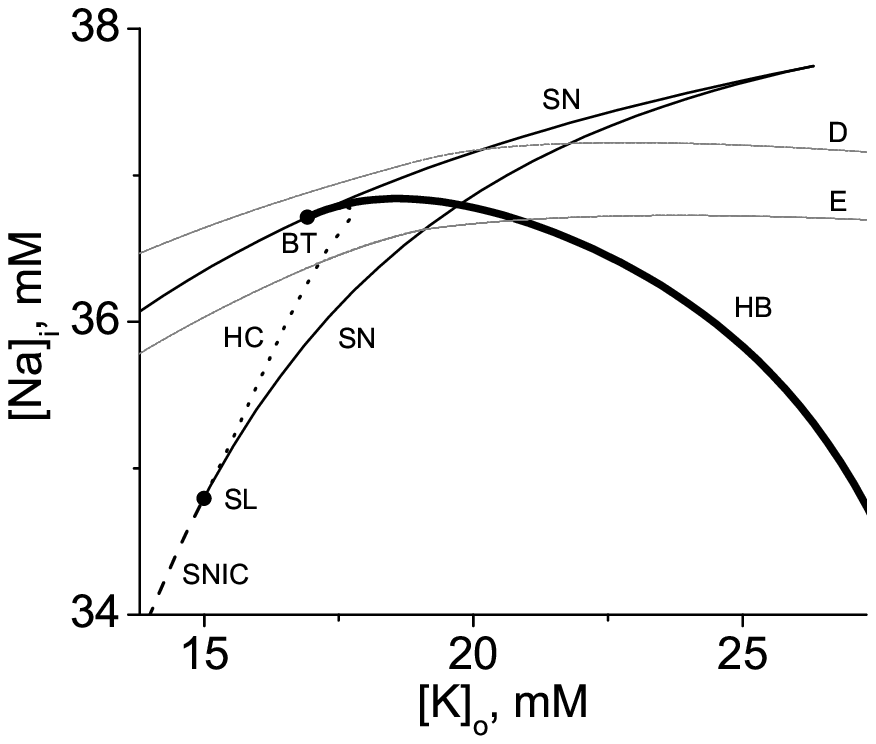}}
\caption{Magnificaton of the upper portion of Figure \ref{SNICHOPF}B. D and E are portions
of ion concentration limit cycles as described in the text.
Abbreviations: SN=saddle-node, BT=Bogdanov-Takens, HC=homoclinic, HB=Hopf, SL=Saddle-node
loop, SNIC=saddle-node infinite-period.}
\label{UpperPart2Dbif}
\end{center}
\end{figure}

Figure \ref{UpperPart2Dbif}
shows a magnification of the upper part of Figure \ref{SNICHOPF}B that reveals more detail.
As the SNIC curve continues up from the lower left corner, a codimension-two bifurcation
known as a saddle-node loop (SL) is encountered \footnote{This bifurcation \cite{GH,Guckenheimer1986}
has also been called a saddle-node homoclinic orbit bifurcation; see \cite{Eugenebook}.}
at approximately  $([K]_o, [Na]_i)=(14.994, 34.795)$ mM.
At this point, the curve splits into two branches corresponding to saddle-node (SN) and homoclinic
bifurcations (HC). The saddle-node branch continues to the upper right and forms a cusp with
another saddle-node branch, while the homoclinic branch continues up, curves around, and terminates
at a (codimension-two) Bogdanov-Takens (BT) point at approximately
$(16.917, 36.714)$ mM.
This point is coincident with the upper SN branch, and is also an endpoint of the Hopf
bifurcation curve (HB).

\begin{figure}
\begin{center}
\scalebox{0.8}{\includegraphics{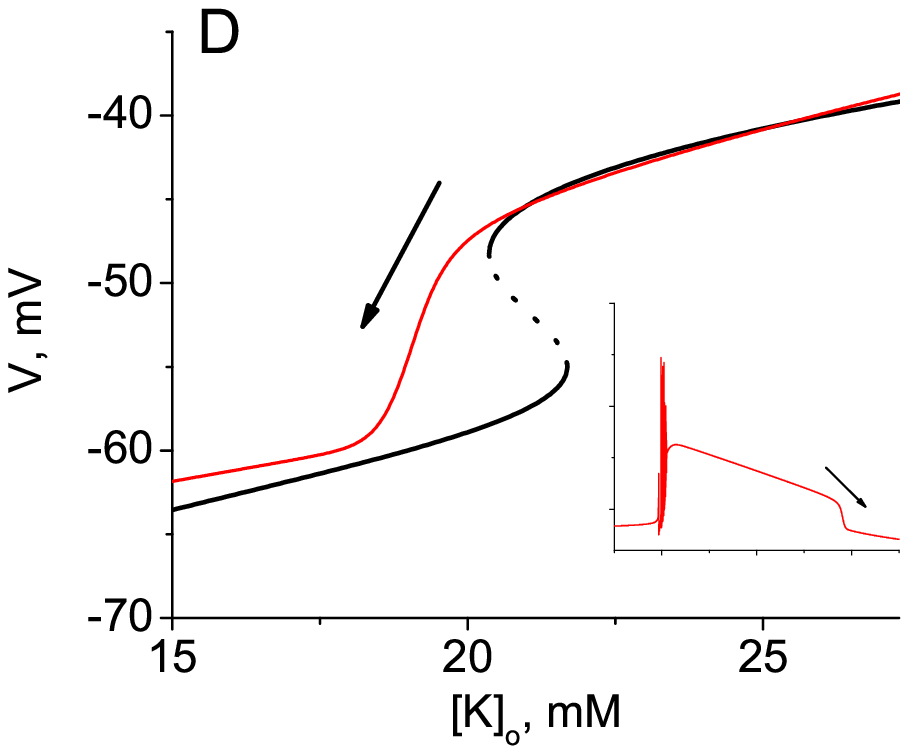}}
\scalebox{0.8}{\includegraphics{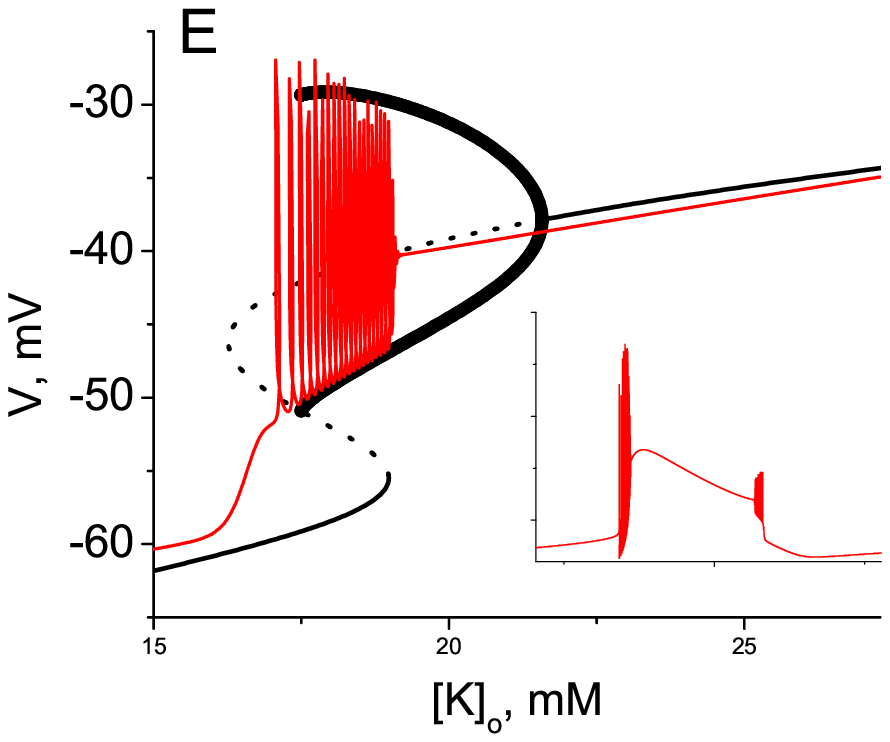}}
\caption{(Color) Diagrams clarifying the burst termination behavior corresponding to ion concentration loops D and E of Figure \ref{UpperPart2Dbif}.
Insets show complete burst time traces of the membrane voltage. In D, $[Na]_i=37.2$ $mM$; in E, $[Na]_i=36.6$ $mM$.}
\label{BurstTermination}
\end{center}
\end{figure}

Superimposed on this diagram is the upper portion of loop D from Figure \ref{Burst_loopmap},
which represents the termination of the burst event.
With the increased magnification, it can be seen that this part of the ion concentration limit cycle
indeed does not cross the Hopf curve. Instead, it crosses the two saddle-node curves.
To clarify the nature of the burst termination, we show in the left panel of Figure \ref{BurstTermination}
(labeled `D') the one-dimensional bifurcation diagram analogous to Figure \ref{SNICHOPF}A for
$[Na]_i=37.2$ mM, along with the burst termination portion of the system's trajectory. Also included
is an inset showing the membrane voltage versus time for one complete burst event.
It can be seen that the termination (arrows) occurs
when the trajectory tracking the upper stable equilibrium branch encounters the second saddle-node
bifurcation and drops to the lower stable equilibrium branch.

The additional detail in Figure \ref{UpperPart2Dbif} permits the identification of a fifth type
of bursting pattern exhibited by our system
(obtained with $\rho=0.9$, $G=10$, $\epsilon=0.5$, $k_{bath}=20$, and $\gamma=0.25$).
E labels the upper portion of an ion concentration
loop that is slightly lower than the D loop. It can be seen that this portion of
loop E crosses, in order, the Hopf, saddle-node, and homoclinic bifurcation curves.
The remaining portion of loop E (not shown) is similar to loop B, and indeed the
membrane voltage traces are similar, but the termination mechanisms are different.
Notably, this type of burst terminates after a few small-amplitude spikes via
homoclinic rather than SNIC bifurcation; see Figure \ref{BurstTermination}E.

Finally, we note that we have also identified parameter sets that give rise to loops that encircle
the BT point in Figure \ref{UpperPart2Dbif} while lying entirely above the SL point,
as well as similar loops occurring upward and to the right of this, including ones
that straddle only the two SN curves. However, the separation of time scales is less clear in these cases.
And although it might exist, we were not able to find a loop that
crosses the SNIC and Hopf curves and then returns above the saddle-node cusp. Such a loop
would exhibit a burst termination that transitions from depolarization block to the rest state smoothly
and without a sudden drop in membrane potential.

\section{Discussion}
In this paper, we have presented a simple model of a single neuron with dynamic
intra- and extra-cellular sodium and potassium concentrations that exhibits periodic
bursting behavior. Our goal has been to present a catalog of the wide variety
of qualitatively different bursting patterns exhibited by this model based on an understanding of
the underlying bifurcation structure. To do this, we have taken advantage of the large
separation of time scales inherent in the model: that of spikes, which occur on the
order of milliseconds, and that of the bursting events, which occur on the order of tens of seconds.
We have focused attention on the nature of the fast spiking dynamics
by freezing the slow ion concentration variables and identifying the asymptotic behavior of the neuron
under these conditions.
This clarifies how oscillations in the slow variables modulate the neuron's
excitability and give rise to bursting. 

A crucial aspect of our model that is typically not present in comparably
simple models (to our knowledge) is the inclusion of sodium concentration dynamics.
Thus, we have {\em two} slow variables ($[Na]_i$ and $[K]_o$) instead of one, and it is therefore possible
for the slow system to undergo bifurcations to oscillatory states. Analysis
complementary to the work presented here, in which bifurcations in the slow dynamics
are analyzed by removing the fast dynamics via averaging, has been presented in \cite{CressmanJCNS}.

For this study, we deliberately chose a single neuron with the classic Hodgkin-Huxley ionic currents in order to focus attention on the
role of the ion concentration dynamics. Consequently, some quantitative aspects of the dynamics (e.g., the value of $[K]_o$ for the onset of depolarization
block in Figure \ref{SNICHOPF}A) are not directly applicable to mammalian systems. More realistic models of mammalian systems would require additional currents and a network
structure with significant synaptic activity. (For example, despite the intriguing similarity of Figure 4D
to phenomena observed in spreading depression, it is known that spreading depression requires
persistent sodium channels \cite{SomjenBook}.) In addition, the absence of synaptic background activity means that
our model neuron transitions abruptly from rest directly to bursting as $[K]_o$ increases. In a more realistic setting, we would
expect to see an intervening regime of tonic spiking due to network effects before the onset of bursting.

Although our approach has concentrated on identifying all possible bursting scenarios
without regard to physiological relevance, we note that bursting morphologies similar
to those that we catalog here have been observed in experimental models of epilepsy,
including a high-potassium model
\cite{JensenYaari}, a high-potassium, low-calcium model \cite{Bikson}, and a
low-magnesium 4-aminopyridine model \cite{Ziburkus}. 
This suggests that despite the obvious limitations of our model, it is capturing
aspects of the essential dynamics in these experiments. And, although there has
been debate about whether a single neuron can undergo a ``seizure" \cite{Connors,Kager2007},
our single-neuron results suggest that ion concentration dynamics may well play an important role
in epilepsy.

It is interesting to note that in
order to obtain our type D burst, in which the burst termination transitions smoothly
from depolarization block back to rest without exhibiting spikes (Figure \ref{Burst_types}D),
we found it necessary to increase the parameter $\gamma$, which is inversely proportional
to the volume of our assumed spherical cell (see the Appendix). With a smaller volume,
the internal sodium concentration is more easily increased to sufficient levels to prevent
spiking. Thus, our model predicts that this type of burst should be most easily observed in
smaller neurons. Accordingly, this behavior has been seen in hippocampal
interneurons (e.g., Figure 4a of \cite{Ziburkus}). Similar behavior in other cases, however, may well be
due to different mechanisms (e.g., calcium \cite{PinskyRinzel} or
chloride accumulation \cite{Rinzel}), and future
work will explore the consequences of including these and other ions \cite{Alexander},
as well as a more realistic collection of channels.

\section{Conclusion}
We have found that a simple model of a single neuron, augmented by
dynamic intra- and extra-cellular ion concentrations, can display
various kinds of periodic bursting behavior. Our work 
emphasizes the importance of ion concentration homeostasis in the maintenance
of the normal physiological state, and suggests that a breakdown in such homeostatic
mechanisms may underlie pathological conditions such as epilepsy.

\section{Acknowledgements}
The authors would like to acknowledge the assistance of Jeremy Owen in creating Figure \ref{DriftingLoops}.
This work was supported by the Collaborative Research in Computational Neuroscience program
at National Institutes of Health via grant R01-MH79502 (EB).

\appendix

\section{Parameters and Equations for Gating Variables}

Parameters not specified above were set as follows:
\[\begin{array}{l}
 {C_m} = 1\;\mu F/c{m^2} \\ 
 {g_{Na}} = 100\;mS/c{m^2} \\ 
 {g_{NaL}} = 0.0175\;mS/c{m^2} \\ 
 {g_K} = 40\;mS/c{m^2} \\ 
 {g_{KL}} = 0.05\;mS/c{m^2} \\ 
 {g_{ClL}} = 0.05\;mS/c{m^2}. \\ 
 \end{array}\]
The sodium activation was instantaneous, with
\[\begin{array}{l}
 {m_\infty }\left( V \right) = \frac{{{\alpha _m}\left( V \right)}}{{{\alpha _m}\left( V \right) + {\beta _m}\left( V \right)}} \\ 
 {\alpha _m}\left( V \right) = \frac{{0.1\left( {V + 30} \right)}}{{1 - \exp \left( { - 0.1\left( {V + 30} \right)} \right)}} \\ 
 {\beta _m}\left( V \right) = 4\exp \left( { - \frac{{V + 55}}{{18}}} \right). \\ 
 \end{array}\]
The remaining gating variables were governed by
 \[\begin{array}{l}
 \frac{{dq}}{{dt}} = \phi \left[ {{\alpha _q}\left( V \right)\left( {1 - q} \right) - {\beta _q}\left( V \right)q} \right],\quad q = h,n \\ 
\end{array}\]
with
\[\begin{array}{l}
 \phi=3.0 \\
 {\alpha _h}\left( V \right) = 0.07\exp \left( { - \frac{{V + 44}}{{20}}} \right) \\ 
 {\beta _h}\left( V \right) = \frac{1}{{1 + \exp \left( { - 0.1\left( {V + 14} \right)} \right)}} \\ 
 {\alpha _n}\left( V \right) = \frac{{0.01\left( {V + 34} \right)}}{{1 - \exp \left( { - 0.1\left( {V + 34} \right)} \right)}} \\ 
 {\beta _n}\left( V \right) = 0.125\exp \left( { - \frac{{V + 44}}{{80}}} \right). \\ 
\end{array}\]

\section{Derivation of Conversion Factors}
To derive the conversion factor $\gamma$, we consider a membrane current density $I$, measured in
$\mu A / cm^2$, and ask: how many ions exit a cell through a patch of
membrane of area $A$ ($cm^2$) in time $\Delta t$ (sec) due to this current? Recalling the convention
that membrane currents are positive-outward, the amount of charge that leaves is
\[\Delta Q = \left( {IA} \right)\Delta t.\]
Assuming monovalent ions, the number of ions is
\[\Delta N = \left( {\frac{{IA}}{q}} \right)\Delta t,\]
where $q = 1.60 \times {10^{ - 19}}$ Coulombs.  Hence,
the rate at which ions pass through the area A is
\[\frac{{dN}}{{dt}} = \left( {\frac{{IA}}{q}} \right).\]
Now assume a spherical cell of volume $V_{in}$ (measured in $cm^3$) with
an intracellular concentration $c_{in}$ of positive ions measured in $ions/cm^3$.
We have
\[{c_{in}} = \frac{N}{{{V_{in}}}},\]
where N is the number of ions within the volume. Hence,
\[\frac{{d{c_{in}}}}{{dt}} = \left( {\frac{1}{{{V_{in}}}}} \right)\frac{{dN}}{{dt}} =  - \frac{1}{{{V_{in}}}}\left( {\frac{{IA}}{q}} \right).\]
The minus sign has been introduced because a positive (outward) current corresponds to a decrease in the number of (positive) ions within the cell.

We wish to express this rate-of-change of concentration in $mM/msec$. The
expression on the right-hand side has units
\[ \frac{{\rm{1}}}{{{\rm{c}}{{\rm{m}}^{\rm{3}}}}}\frac{{\frac{{{\rm{\mu A}}}}{{{\rm{c}}{{\rm{m}}^{\rm{2}}}}}{\rm{c}}{{\rm{m}}^{\rm{2}}}}}{{{\rm{Coul}}}} \to {10^{ - 6}}\frac{{{\rm{ions}}}}{{{\rm{c}}{{\rm{m}}^{\rm{3}}}{\rm{sec}}}} \to \frac{1}{{{{10}^3}{N_A}}}\frac{{{\rm{millimole}}}}{{{\rm{msec}}\;{\rm{L}}}}, \]
where $N_A = 6.02 \times {10^{23}}$ is Avogadro's number. Since $1$ $millimole/L=$ $1$ $mM$, we have
\[\frac{{d{c_{in}}}}{{dt}} =  - \left( {\frac{1}{{{{10}^3}}}} \right)\frac{{IA}}{{{N_Aq}{V_{in}}}}.\]
We write
\begin{equation}
\frac{{d{c_{in}}}}{{dt}} =  - \left( {\frac{1}{{{{10}^3}}}} \right)\gamma I\
\label{Naconc}
\end{equation}
with
\[\gamma  \equiv \frac{A}{{{N_A}q{V_{in}}}} = \frac{A}{{F{V_{in}}}},\]
where $F = {N_A}q$ is the Faraday constant. Equation (\ref{Naconc}) is used
for the intracellular sodium in Equation (\ref{concdiffeqs}).

Now assume that the sphere has radius $r$, and let $A$ be its total surface area.
(More correct would be to let $A$ be just the area of the channel pores.) Then,
\[\gamma  =  \frac{3}{{Fr}}.\]
Using $r=7$ $\mu m$, we obtain $\gamma=4.45 \times {10^{ - 2}}$. Conversely, for
$\gamma=0.25$ and $1.0$, we obtain $r=1.24$ and $0.31$ $\mu m$, respectively.

Consider now the concentration of positive ions in the extracellular space. The volume in the expression
for $\gamma$ must be replaced with the extracellular volume, and the negative sign in Equation \ref{Naconc}
must be removed. Let $\beta  \equiv {V_{in}}/{V_{out}}$. Then
$\gamma \to \gamma \beta$, and we have
\begin{equation}
\frac{{d{{c}_{out}}}}{{dt}} = \left( {\frac{1}{{{{10}^3}}}} \right)\gamma \beta I.
\label{Kconc}
\end{equation}
Equation (\ref{Kconc}) is used for the extracellular potassium in Equation (\ref{concdiffeqs}).


\begin{thebibliography}{}

\bibitem{HH}
Hodgkin, A.L., Huxley, A.F.:
A quantitative description of membrane current and its application to conduction and excitation in nerve.
J. Physiol. 117(4), 500–544 (1952)

\bibitem{FH}
Frankenhaeuser, B., Hodgkin, A.L.: The after-effects of impulses in the giant nerve fibres of {\it Loligo}.
J. Physiol. 131, 341-376 (1956)

\bibitem{Grafstein}
Grafstein, B.:
Mechanism of spreading cortical depression.
J. Neurophysiol 19, 154–171 (1956)

\bibitem{Green}
Green, J.D.:
The Hippocampus.
Physiolgical Reviews 44, 561-608 (1964)

\bibitem{FR}
Fertziger, A.P., Ranck, J.B. Jr.:
Potassium accumulation in interstitial space during epileptiform seizures.
Experimental Neurology 26, 571-85 (1970)

\bibitem{FrohlichReview}
Fr{\" o}hlich, F., Bazhenov, M., Iragui-Madoz, V., Sejnowski, T.J.:
Potassium dynamics in the epileptic cortex: new insights on an old topic.
The Neuroscientist 14(5), 422-433 (2008)

\bibitem{Kager2002}
Kager, H., Wadman, W.J., Somjen, G.G.:
Conditions for the triggering of spreading depression studied with computer simulations.
J. Neurophysiol. 88, 2700-2712 (2002)

\bibitem{Somjen2008b}
Somjen, G.G., Kager, H., Wadman, W.J.:
Calcium sensitive non-selective cation current promotes seizure-like discharges and spreading depression in a model neuron.
J. Comp. Neurosci. 26, 139-147 (2008)

\bibitem{Bazhenov2004}
Bazhenov, M., Timofeev, I., Steriade, M., Sejnowski, T.J.:
Potassium model for slow (2-3 Hz) in vivo neocortical paroxysmal oscillations.
J. Neurophysiol. 92, 1116-1132 (2004)

\bibitem{Frohlich2006}
Fr{\" o}hlich, F., Bazhenov, M., Timofeev, I., Steriade, M., Sejnowski, T.J.:
Slow state transitions of sustained neural oscillations by activity-dependent modulation of intrinsic excitability.
Journal of Neuroscience 26(23), 6153-6162 (2006)

\bibitem{Park}
Park, E.H., Durand, D.M.:
Role of potassium lateral diffusion in non-synaptic epilepsy: a computational study.
Journal of Theoretical Biology 238, 666–682 (2006)

\bibitem{Kager2007}
Kager, H., Wadman, W.J., Somjen, G.G.:
Seizure-like afterdischarges simulated in a model neuron.
J. Comp. Neurosci. 22, 105-128 (2007)

\bibitem{Somjen2008a}
Somjen, G.G., Kager, H., Wadman, W.J.:
Computer simulations of neuron-glia interactions mediated by ion flux.
J. Comp. Neurosci. 25, 349-365 (2008)

\bibitem{Postnov2009}
Postnov, D.E., M{\" u}ller, F., Schuppner, R.B., Schimansky-Geier, L.:
Dynamical structures in binary media of potassium-driven neurons.
Physical Review E 80, 031921 (2009)

\bibitem{Frohlich2010}
Fr{\" o}hlich, F., Sejnowski, T.J., Bazhenov, M.:
Network bistability mediates spontaneous transitions between normal and pathological brain states.
Journal of Neuroscience 30(32), 10734 –10743 (2010)

\bibitem{Kager2000}
Kager, H., Wadman, W.J., Somjen, G.G.:
Simulated seizures and spreading depression in a neuron model incorporating interstitial space and ion concentrations.
Journal of Neurophysiology 84, 495–512 (2000)

\bibitem{SomjenBook}
Somjen, G.G.: Ions in the Brain. Oxford University Press, New York (2004)

\bibitem{CressmanJCNS}
Cressman, J.R., Ullah, G., Schiff, S.J., Barreto, E.:
The Influence of Sodium and Potassium Dynamics on Excitability, Seizures, and the Stability of Persistent States: I. Single Neuron Dynamics.
Journal of Computational Neuroscience 26, 159-170 (2009)

\bibitem{DayanAbbott}
Dayan, P., Abbott, L.F.: Theoretical Neuroscience. MIT Press, Cambridge (2001)

\bibitem{Eugenebook}
Izhikevich E.:
Dynamical systems in Neuroscience.
MIT Press, Cambridge (2007)

\bibitem{SNIPERref}
Martens, E., Barreto, E., Strogatz, S.H., Ott, E., So, P., Antonsen, T.M.:
Exact results for the Kuramoto model with a bimodal frequency distribution.
Physical Review E 79, 026204 (2009)

\bibitem{Bikson}
Bikson, M., Hahn, P.J., Fox, J.E., Jefferys, J.G.R.:
Depolarization block of neurons during maintenance of electrographic seizures.
J. Neurophysiol. 90, 2402-2408 (2003)

\bibitem{Cymbalyuk1}
Shilnikov, A.L., Calabrese R., and Cymbalyuk, G.S.:
Mechanism of bi-stability: tonic spiking and bursting in a neuron model.
Physical Review E 71, 056214 (2005) 

\bibitem{Cymbalyuk2}
Cymbalyuk, G.S., Calabrese R., and Shilnikov, A.L.:
How a neuron model can demonstrate co-existence of tonic spiking and bursting?
Neurocomputing 65-66: 869-875 (2005)

\bibitem{FrohlichBazhenov2006}
Fr{\" o}hlich, F., Bazhenov, M.:
Coexistence of tonic firing and bursting in cortical neurons.
Physical Rev E 74, 031922 (2006)

\bibitem{GH}
Guckenheimer, J., Holmes, P.:
Nonlinear Oscillations, Dynamical Systems, and Bifurcations of Vector Fields.
Springer-Verlag, New York (1983)

\bibitem{Guckenheimer1986}
Guckenheimer J.:
Multiple bifurcation problems for chemical reactors.
Physica D 20(1), 1-20 (1986)

\bibitem{JensenYaari}
Jensen, M.S., Yaari, Y.:
Role of intrinsic burst firing, potassium accumulation, and electrical coupling in the elevated potassium model of hippocampal epilepsy.
J. Neurophysiol. 77, 1224-1233 (1997) 

\bibitem{Ziburkus}
Ziburkus, J., Cressman, J.R., Barreto, E., Schiff, S.J.:
Interneuron and pyramidal cell interplay during in vitro seizure-like events.
J. Neurophysiol. 95, 3948–3954 (2006)

\bibitem{Connors}
Connors, B.W., Telfeian, A.E.:
Dynamic properties of cells, synapses, circuits and seizures in neocortex.
In: Neocortical Epilepsies, Williamson, P.D. et al. (eds.) Advances in Neurology, 84:141-152, 2000.

\bibitem{PinskyRinzel}
Pinsky, P.F., Rinzel, J.:
Intrinsic and network rhythmogenesis in a reduced Traub model for CA3 neurons.
Journal of Computational Neuroscience 1, 39-60 (1994)

\bibitem{Rinzel}
Marchetti, C., Tabak, J., Chub, N., O'Donovan, M.J., Rinzel, J.:
Modeling spontaneous activity in the developing spinal cord using activity-dependent variations of intracellular chloride.
Journal of Neuroscience 25(14), 3601-3612 (2005)

\bibitem{Alexander}
Komendantov, A., Cressman, J.R., Barreto, E.:
Ion concentration homeostasis and the regulation of neuronal firing activity: the role of cation-chloride cotransporters.
BMC Neuroscience 2010, 11(Suppl 1), P27 (2010).

\end{thebibliography}
\end{document}